# *In Situ* Thermal Decomposition of Exfoliated Two-Dimensional Black Phosphorus

Xiaolong Liu[1†], Joshua D. Wood[2†], Kan-Sheng Chen[2], EunKyung Cho[2], and Mark C. Hersam[1,2,3,4*]

*[1]Graduate Program in Applied Physics, Northwestern University, Evanston, IL 60208, USA*
*[2]Department of Materials Science and Engineering, Northwestern University, Evanston, IL 60208, USA*
*[3]Department of Chemistry, Northwestern University, Evanston, IL 60208, USA*
*[4]Department of Medicine, Northwestern University, Evanston, IL 60208, USA*

[*] Correspondence should be addressed to m-hersam@northwestern.edu

[†] These authors contributed equally.

**Abstract**

With a semiconducting band gap and high charge carrier mobility, two-dimensional (2D) black phosphorus (BP) – often referred to as phosphorene – holds significant promise for next generation electronics and optoelectronics. However, as a 2D material, it possesses a higher surface area to volume ratio than bulk BP, suggesting that its chemical and thermal stability will be modified. Herein, an atomic-scale microscopic and spectroscopic study is performed to characterize the thermal degradation of mechanically exfoliated 2D BP. From *in situ* scanning/transmission electron microscopy, decomposition of 2D BP is observed to occur at ~400 °C in vacuum, in contrast to the 550 °C bulk BP sublimation temperature. This decomposition initiates *via* eye-shaped cracks along the [001] direction and then continues until only a thin, amorphous red phosphorous like skeleton remains. *In situ* electron energy loss spectroscopy, energy-dispersive X-ray spectroscopy, and energy-loss near-edge structure changes provide quantitative insight into this chemical transformation process.

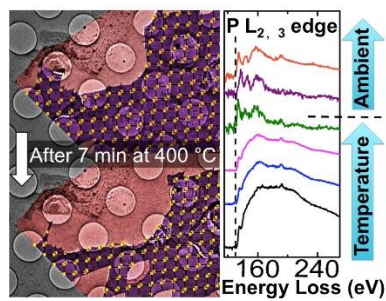

**TOC Figure**

**Manuscript**

Two-dimensional (2D) nanomaterials including graphene,[1] hexagonal boron nitride (h-BN),[2] $MoS_2$,[3-4] $sp^2/sp^3$ silicon,[5] and black phosphorus (BP)[6-25] have attracted significant interest in electronics and photonics applications. Since these materials possess high surface area to volume ratios, their surface chemistries need to be carefully considered when developing device processing and fabrication methods. Towards this end, previous work has found that graphene and h-BN are chemically and thermally robust,[1-2] while 2D $MoS_2$ undergoes chemical transformations during chemical exfoliation[26] and high temperature annealing.[4] In contrast, as one of the newest members of the 2D nanomaterial portfolio,[13, 15, 27] little is known about the chemical and thermal stability of 2D BP.

Although bulk BP is the most thermodynamically stable phosphorus allotrope, it can suffer from electrochemical[28] and ambient oxidation.[23, 29-31] Since nanomaterials often show lower decomposition temperatures (e.g. carbon nanotubes,[32] graphene nanoribbons,[33] and Al nanoparticles[34]) compared to their bulk counterparts, similar effects can be expected for 2D BP. Furthermore, even for bulk BP, there is little consensus on basic thermal stability parameters. For example, the reported melting temperature of bulk BP varies from 600 °C[35-37] to >1000 °C.[38] Therefore, it is important to clarify thermal phenomena in 2D BP before it can be effectively employed in applications.

Herein, we quantitatively assess the thermal decomposition of mechanically exfoliated BP. In particular, *in situ* scanning/transmission electron microscopy (S/TEM) is performed on BP flakes transferred onto TEM grids, allowing for chemical, morphological, and crystallographic decomposition intermediates to be determined by selected area diffraction (SAD), electron energy loss spectroscopy (EELS), energy-dispersive X-ray spectroscopy (EDS), and X-ray photoelectron spectroscopy (XPS). These morphological and spectroscopic measurements reveal that the *in situ* sublimation temperature of exfoliated 2D BP is ~400 °C with the resulting eye-shaped defects proceeding along the [001] direction and ultimately culminating in a red phosphorus like skeleton.

Fig. 1(a) reveals a bright field TEM image for an exfoliated BP flake, where the orthorhombic crystalline character is confirmed in the inset SAD pattern. Fig. 1(b) is a high-magnification image of BP showing lattice fringes. To minimize the influence of the electron beam while maintaining spatial resolution, a 120 to 200 keV acceleration voltage is used, and the sample is moved into the beam only when imaging (see Supporting Information for more details). The BP flakes are stored in dry $N_2$ until TEM is performed. Figs. 1(a, b) indicate that the BP flakes are crystalline with no entrapped species[29] or ambient degradation evident.[23, 28, 30] Atomic force microscopy (AFM) measurements also confirm that the sample preparation method in this study does not cause flake degradation (Fig. S1). Comparatively, BP flakes exposed to ambient conditions for several days show entrapped species (Fig. S2). Figs. 1(c-g) show the edge of a BP flake at different annealing conditions with the insets providing the corresponding SAD patterns. Compared with the unannealed case (Fig. 1(c)), the wrinkles in the BP disappear, with the flake becoming flatter after 20 min of 200 °C annealing (Fig. 1(d)). The flake appears unchanged after further 300 °C annealing for 20 min (Fig. 1(e)) with SAD patterns showing unchanged BP crystal orientation. After annealing at 400°C for 20 min (Fig. 1(f)), irregular patterns appear, and the SAD pattern indicates BP amorphization. The flake remains amorphous after 500 °C annealing for 20 min (Fig. 1(g)). EDS spectra for the flake before (Fig. 1(h)) and after (Fig. 1(i)) annealing exhibit a P Kα peak at 2 keV that decreases in intensity after annealing. This decrease can be attributed to the lower interaction volume in thinner BP flakes, as detailed below.

Fig. 2 considers the dynamics of BP amorphization at 400 °C, using low magnification images of the flake from Fig. 1. Figs. 2(a-c) highlight the invariance of the flake to heating up to 300 °C. However, when the flake is heated at 400 °C for 5 min, the flake edge retracts, leaving a thin layer behind (Fig. 2(d)). Additional BP decomposition proceeds from the edges until 20 min exposure (Figs. 2(e-g)), after which a thin skeleton predominantly remains. This thin skeleton withstands an additional 20 min of annealing at 500 °C (Fig. 2(h)) where it remains continuous over the holes of the Quantifoil TEM support. The SAD patterns and EDS spectra in Figs. 1(f-i) reveal that the skeleton is an amorphous phosphorus structure. Since a phosphorus skeleton remains and no liquid-like features appear during decomposition, the BP degradation in Figs. 2(d-g) appears to proceed by BP sublimation and not melting. Since the BP flakes transferred onto the TEM grid are 10 to 40 nm thick, as seen in AFM and confirmed by calculations based on TEM contrast[39] (Fig. S3), their suppressed thermal decomposition temperature compared to the melting point of bulk BP is consistent with other low dimensional nanomaterials with high surface area to volume ratios.[40-42]

BP sublimation occurs at flake edges and defects, and then propagates as eye-shaped cracks. Figs. 3(a-c) give higher magnification images of this crack evolution as the flake is heated to 400

°C for 5 min, 8 min, and 12 min, respectively. In Fig. 3(d), a SAD pattern for this flake (taken at 300 °C) indicates that all cracks are along the [001] direction. Similar to the sublimation of graphene,[43-44] cracks grow larger and coalesce, but, unlike graphene, the BP cracks maintain a regular shape during sublimation. As such, the crystallographic structure of BP appears important in the decomposition mechanism. We first consider the effects of the *a-c* plane (hereafter "in-plane") thermal expansion as the source of the oriented cracks. While relatively few and conflicting reports on the coefficients of thermal expansion (CTE) for BP exist,[38, 45] we can estimate CTE values for the *a* and *c* lattice parameters using our SAD patterns. Figs. 3(e-f) examine six BP flakes and determine average *a* and *c* values from the SAD data. The error bar for each data point consists of standard deviation and TEM system error, estimated to be ±1%.[46] Bulk BP lattice parameters at elevated temperatures are calculated based on bulk BP CTE from the literature.[38] We find that the CTE values of exfoliated BP flakes along the [100] and [001] directions are $\alpha_a = (90.3 \pm 6.4) \times 10^{-6}$/°C and $\alpha_c = (93.2 \pm 12.7) \times 10^{-6}$/°C, three and two fold higher than reported values for bulk BP,[38] respectively. Since the two in-plane CTE values for exfoliated BP flakes are nearly identical, thermal expansion does not appear to be the main cause for the anisotropic, eye-shaped cracks.

However, the anisotropic, buckled atomic structure (Fig. 3(g)) of BP suggests a mechanism for the eye-shaped crack formation. For bulk BP, each phosphorus atom has three single bonds (see Fig. S4), with two in-plane bonds and a third out-of-plane bond.[47] Assuming that sublimation begins with a vacancy defect,[44] denoted by a green circle in Fig. 3(h), BP decomposition then continues by removal of P atoms along the perimeter. The perimeter P atoms surrounding the growing crack fall into several categories including P atoms with only one bond to the surrounding crystal and P atoms with two bonds to the surrounding crystal. The second category can be further subdivided into P atoms with two in-plane bonds and P atoms with one in-plane bond and one out-of-plane bond. Assuming that P atoms with only one bond desorb first followed by P atoms with two in-plane bonds and finally P atoms with one in-plane bond and one out-of-plane bond, then the crack will evolve in an anisotropic manner that is consistent with the observed eye-shape along the [001] direction. Snapshots of this proposed sublimation model are shown in Figs. 3(h-j) with further details included in the Supporting Information (Fig. S5). This model works with arbitrary initial crack directions and irregular vacancy geometries (Fig. S6). For BP multilayers, formation of cracks in one layer will expose the next inner layer, which initiates sublimation at the same point. Hence, sublimation of BP multilayers will also exhibit eye-shaped cracks as observed experimentally in Fig. 3.

Concurrent *in situ* STEM heating and EELS measurements provide chemical information for the BP decomposition process. Figs. 4(a-g) contain the transmission electron (TE) images for a range of annealing conditions. After 20 min annealing at 500 °C, the sample is cooled *in situ* to 37 °C, upon which it is exposed to ambient conditions (~26 °C, relative humidity ~38%) for 5 min. The dark contrast in the TE images in Fig. 4(h) implies that the flake has become thicker after ambient exposure. In addition, the flake is invariant to additional ambient treatment, as evidenced by the 24 hr ambient exposure preceding the image in Fig. 4(i). Fig. 4(j) shows the EELS spectra taken at different annealing conditions for the extracted and normalized P $L_{2,3}$ edge at 132 eV. The EELS spectra for the O K edge at 532 eV are provided in Fig. S7.

To better highlight the change in the energy-loss near-edge structure (ELNES), the derivatives of the EELS P $L_{2,3}$ edge are presented in Fig. 4(j). The existence of the P $L_{2,3}$ edge after 400°C annealing indicates that the thin skeleton contains P, in agreement with EDS spectrum shown in Fig. 1(i). The ELNES of the P $L_{2,3}$ is invariant before 400°C annealing and changes its line shape qualitatively at 400°C, denoting a different P bonding state.[48-50] The P $L_{2,3}$ ELNES modifications after ambient exposure are negligible, implying a similar P oxidation state before and after ambient. Although the O K edge exists before annealing from the BP preparation process, it is more evident after ambient exposure (Fig. S7). Similarly, XPS data taken after ambient (Fig. S8) shows oxygenated P.[29, 51] The TEM images in Figs. S9, S10 and S11 further highlight this BP chemical decomposition.

A previous study[36] showed that heating bulk BP to 777 °C at ~0.5 GPa pressure produces a 0.3-0.5 mm thick layer of red P on the BP crystal surface. Brazhkin *et al.* also produced a phase diagram for bulk black and red P, showing a phase transition between the black and red phases at 597 °C. This solid red P exists until 620 °C without applied pressure. In general, red P is a highly reactive, amorphous structure. These characteristics suggest that the thin skeleton remaining after BP decomposition is related to red P. Furthermore, previous work has shown that BP degrades in ambient conditions.[23, 29] Oxygenated $H_2O$ has been suggested as a source for BP ambient degradation,[29] and red P is even more hygroscopic than BP. The rapid gettering of oxygenated $H_2O$ by red P following ambient exposure is a plausible explanation for the observed increase in thickness observed in Figs. 4(h, i).

In summary, we have employed a suite of atomically precise microscopy and spectroscopy techniques to characterize the *in situ* thermal decomposition of exfoliated BP at ~400 °C. The high surface area and nanoscale thickness of exfoliated BP likely explains the lower decomposition temperature relative to bulk BP. Sublimation is initiated with the formation of eye-shaped cracks along the [001] direction in a manner consistent with the anisotropic in-plane atomic structure of BP. After thermal decomposition, an amorphous red P like skeleton persists, as inferred from ambient exposure, SAD, EELS, and XPS data. Overall, this study provides insight into the thermal limits of 2D BP, thus facilitating the development of suitable processing methods for BP-based devices.

## Acknowledgments

This research was supported by the Office of Naval Research (N00014-14-1-0669) and the W. M. Keck Foundation. The work made use of the NUANCE Center, which has received support from the MRSEC (NSF DMR-1121262), State of Illinois, and Northwestern University. The authors kindly thank Dr. Jinsong Wu, Dr. Shuyou Li, Dr. Langli Luo, and Dr. Fengyuan Shi for their help with TEM measurements.

## Supporting Information

The supporting information contains methods and additional experimental data. This document is available free of charge via the Internet at http://pubs.acs.org.

## Author Contributions

†These authors (X.L. and J.D.W.) contributed equally to this work.

## Conflict of Interest

The authors declare no competing financial interests.

**Figures**

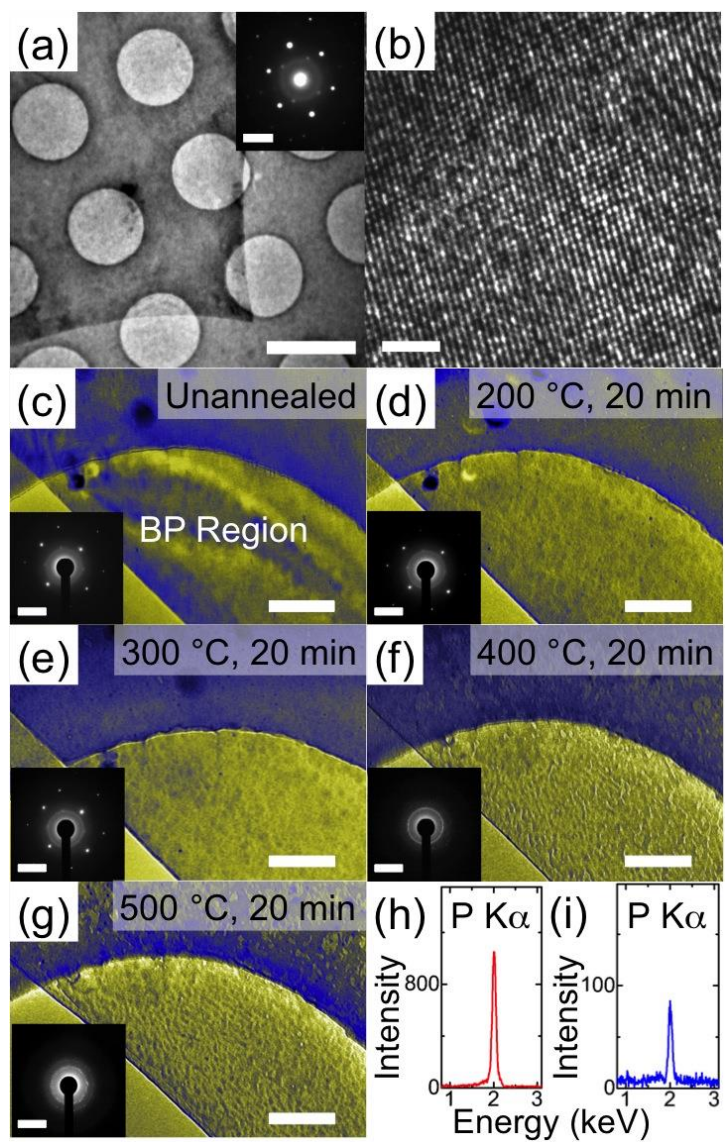

**Figure 1.** TEM images of exfoliated BP and *in situ* heating of BP. **(a)** Low-magnification TEM bright field image of a BP flake. Scale bar is 2 µm. *Inset:* Selected area diffraction (SAD) pattern of the flake. Scale bar is 5 nm$^{-1}$. **(b)** High-magnification TEM bright field image of BP showing lattice fringes. Scale bar is 2 nm. **(c-g)** TEM bright field images (false colored) of a thin BP flake after **(c)** introduction to the TEM and heated at **(d)** 200 °C, **(e)** 300 °C, **(f)** 400 °C, and **(g)** 500 °C, respectively. All heating times are 20 min long. Scale bars are 200 nm. *Insets:* SAD patterns for each heating stage. Scale bars are 5 nm$^{-1}$. Irregular patterns and amorphization occur past 400 °C. EDS spectra of BP flake **(h)** before and **(i)** after 400 °C *in situ* heating. In both cases, a P Kα peak is apparent.

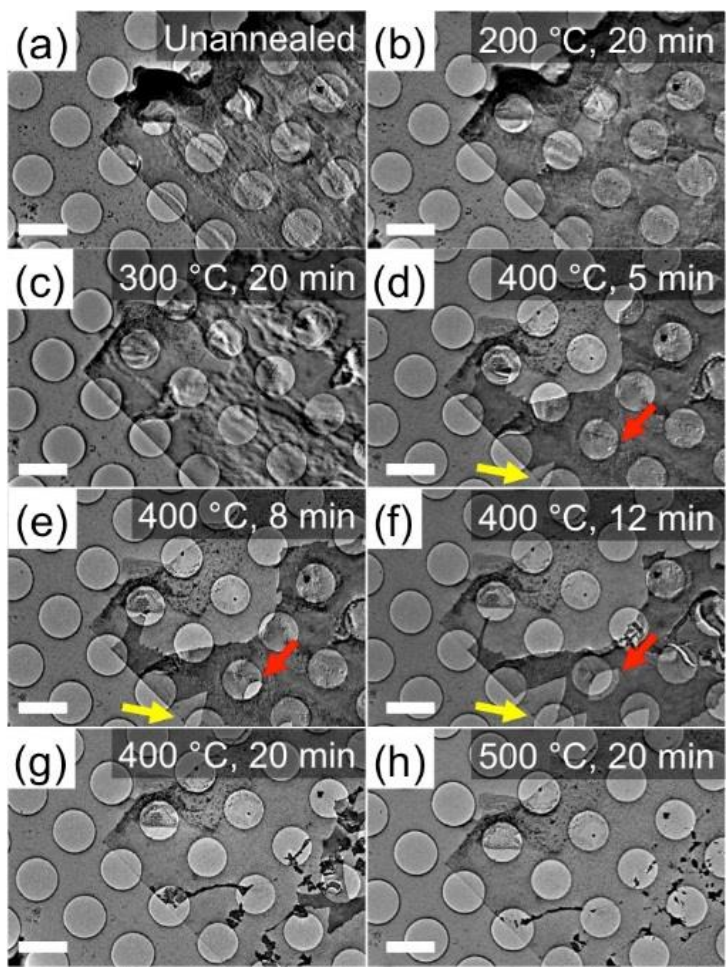

**Figure 2.** Dynamics of *in situ* BP sublimation. Low-magnification TEM images for annealing conditions of: (**a**) unannealed, (**b**) 20 min at 200 °C, (**c**) 20 min at 300 °C, (**d**) 5 min at 400 °C, (**e**) 8 min at 400 °C, (**f**) 12 min at 400 °C, (**g**) 20 min at 400 °C, (**h**) 20 min at 500 °C, respectively. When the flake is heated at 400 °C for 5 min, the flake begins to decompose, as denoted by the yellow and red arrows in (d-f). The degradation continues until 500 °C, after which a thin skeleton remains. Scale bars are 2 μm.

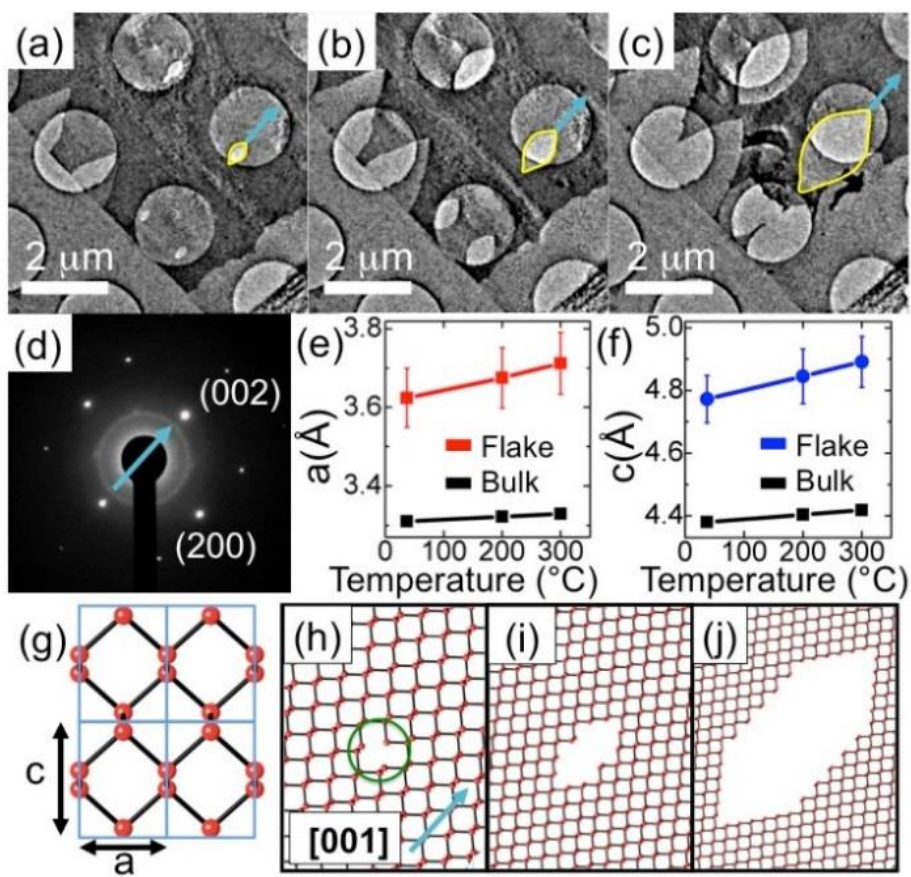

**Figure 3.** Formation of eye-shaped cracks in BP during sublimation. **(a-c)** Zoomed-in image of the flake in Fig. 2, heated for 5 min, 8 min, and 12 min at 400 °C, respectively. Eye-shaped cracks (yellow) form and grow. The blue arrow indicates the propagation direction. **(d)** SAD pattern of the flake at 300 °C, showing the [001] crack propagation direction. **(e, f)** Temperature-induced increase in BP lattice parameters $a$ and $c$. Bulk values from Madelung.[38] **(g)** In-plane lattice schematic for BP. **(h-j)** Snapshots from a BP sublimation model, describing the formation of eye-shaped cracks along the [001] direction (blue arrow).

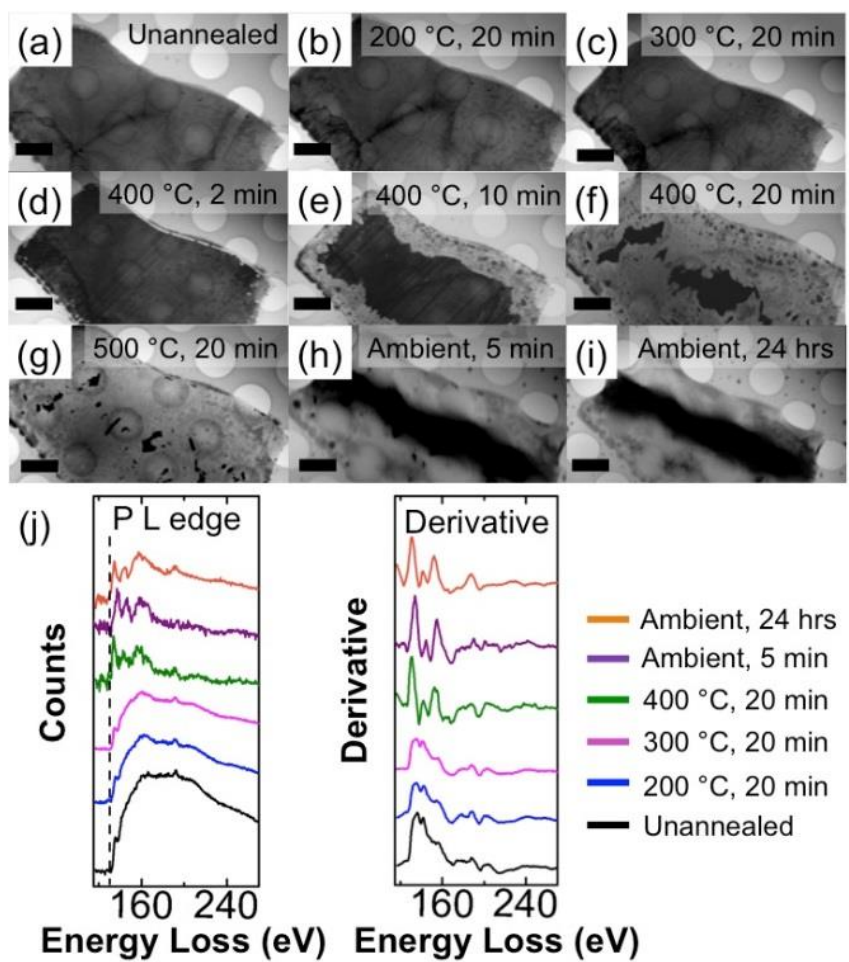

**Figure 4.** *In situ* STEM heating of BP and corresponding EELS spectra. STEM transmitted electron (TE) images of BP after **(a)** introduction to the STEM and heated at **(b)** 200 °C for 20 min, **(c)** 300 °C for 20 min, **(d)** 400 °C for 2 min, **(e)** 400 °C for 10 min, **(f)** 400 °C for 20 min, **(g)** 500 °C for 20 min. Consistent with *in situ* TEM heating, BP sublimation begins at ~400 °C. **(h, i)** STEM TE image of BP exposed to ambient for 5 min and 24 hr after *in situ* heating, respectively. Scale bars are 2 µm. **(j)** P L edge and derivative EELS spectra during each stage of *in situ* heating. P $L_{2,3}$ ELNES line shapes qualitatively change after 400 °C.

# <u>Supporting Information</u>

# *In Situ* Thermal Decomposition of Exfoliated

# Two-Dimensional Black Phosphorus

Xiaolong Liu[1†], Joshua D. Wood[2†], Kan-Sheng Chen[2], EunKyung Cho[2], and

Mark C. Hersam[1,2,3,4*]

[1]*Graduate Program in Applied Physics, Northwestern University, Evanston, IL 60208, USA*
[2]*Department of Materials Science and Engineering, Northwestern University, Evanston, IL 60208, USA*
[3]*Department of Chemistry, Northwestern University, Evanston, IL 60208, USA*
[4]*Department of Medicine, Northwestern University, Evanston, IL 60208, USA*

[*] Correspondence should be addressed to <u>m-hersam@northwestern.edu</u>
† These authors contributed equally.

**Contents:**

## Experimental Details

*Sample Preparation*

Black phosphorus (BP) crystals with purity ~99.998% were purchased from Smart Elements and stored them in a $N_2$ glove box. To make TEM samples, BP was exfoliated with Scotch tape onto 300 nm $SiO_2$ on degenerately doped Si. Immediately after exfoliation, 950 A4 poly(methyl methacrylate) (PMMA) purchased from MicroChem was spin coated onto BP flakes (twice, 2000 RPM, 1 min each time), and the anisole solvent was then baked out at ~150 °C for 5 min. To etch $SiO_2$/Si and free PMMA/BP, the PMMA/BP/$SiO_2$/Si stack was soaked in 2 M KOH. This method is also used in a previous BP study.[1] Commercially, alkaline solutions are used to dissolve white phosphorus in red phosphorus.[2] Due to the higher stability of BP than red phosphorus, we expect KOH will not significantly affect BP. To remove etching residues, the PMMA/BP films were rinsed in ultrapure deionized (DI) water (18.2 MΩ·cm). Using self-locking tweezers, the films were picked up with Quantifoil TEM grids (2 μm holes, 200 mesh Au grid, Ted Pella, Inc.). Entrapped water was removed from the interface of the PMMA/BP films and the TEM grids by successive ~70 °C and ~150 °C heating, each for 10 min. The PMMA was then dissolved with hot acetone vapor (~45 to 55 °C) in necked Erlenmeyer flasks. To mitigate BP degradation in ambient conditions, the grids were stored in a $N_2$ glove box before all measurements. For the AFM control sample (Fig. S1), the same exfoliation and transfer methods were used except that the flakes were transferred onto a Si/$SiO_2$ wafer.

*Scanning/Transmission Electron Microscopy (S/TEM)*

SAD imaging, and low and high magnification TEM characterization (Figs. 1(a, b) in the main manuscript) of BP were performed using a JEOL JEM-2100 TEM at an accelerating voltage of 200 keV. To minimize beam damage, high beam current was only used when taking high magnification images and selected area diffraction (SAD) patterns. Under high-magnification conditions (Fig. 1(b)) at 200 keV acceleration voltage, electron beam-induced amorphization was not significant if the exposure times were short (< 30 s). Conversely, for low-magnification imaging, beam-induced amorphization was negligible. Low beam current was used to navigate and locate BP flakes. *In situ* TEM heating was performed with a Hitachi HT-7700 Biological TEM with an accelerating voltage of 120 keV. SAD patterns were calibrated against a diffraction standard sample (evaporated aluminum, from Ted Pella) at the same accelerating voltage (120 kV) and nominal camera length (*ca.* 0.3 m). *In situ* STEM heating was performed with a Hitachi HD-2300 Dual EDS Cryo STEM with an accelerating voltage of 200 keV. EELS spectra were collected with a 50 μm aperture with convergent semi-angle of 4 mrad and a collection semi-angle of 7 mrad. The typical pressure in the S/TEM columns was ~$1\times10^{-7}$ Torr. Ambient exposure of S/TEM samples before introducing into high vacuum was ~5 min, at a relative humidity of $42.8 \pm 1.9\%$.

*In situ* heating was performed with a Gatan 672 single tilt heating holder (resistive heating) and SmartSet Model 901 hot stage controller. The sample sat in a tantalum heat source with good thermal contacts. The specimen temperature was measured using a Pt/Pt(13%)Rh(87%) thermocouple, which was spot welded to the heat source body. The sample temperature may be slightly lower than the thermocouple reading due to radiative losses between the heat source and the sample. At room temperature in ambient conditions, the temperature reading from the controller was 22.5 °C and the readout from a thermometer (EXTECH 445715) was 22.7 °C. Additionally, the electron beam and TEM/STEM column only heated the sample to ~37 °C preceding intentional annealing.

*X-ray Photoelectron Spectroscopy (XPS)*

XPS spectra were taken at a base pressure of ~$5\times10^{-10}$ Torr on a Thermo Scientific ESCALAB 250 Xi using a monochromated Al $K_\alpha$ source at 1486.7 eV. The binding energy resolution was 0.1 eV, the spot size was 400 μm, and a charge compensating flood gun was utilized for the BP on TEM grid samples. Five scans were averaged for each core level spectrum, and the data were collected at a dwell time of 100 ms and a pass energy of 15 eV. Regardless of the flood gun charge compensation, all spectra were corrected to adventitious carbon at 284.8 eV. On the software suite Avantage (Thermo Scientific), all subpeaks were fit with a modified Shirley background,[3] known as a "Smart" background in the software. For a particular core level, each subpeak shared the same amount of Gaussian-Lorentzian (GL) mixing. If GL mixing became 0% (Lorentzian) or 100% (Gaussian), then the subpeaks were locked at 30% GL mixing (as is typically used in CasaXPS and other software suites). All subpeaks with full-width at half-maximum (FWHM) greater than 0.5 eV were disregarded. Doublets were used for the p core levels for phosphorus and silicon. Subpeaks were added until the residual level was minimized.

*Atomic Force Microscopy (AFM)*

All AFM measurements were taken in air with ~300 kHz cantilevers on an Asylum Cypher AFM in tapping mode. Images were taken with a pixel resolution of 512×512 at a scanning rate of ~0.5 Hz. The BP TEM sample was mounted onto a metal specimen disc with BP flakes facing the cantilever. Two small pieces of adhesive tape were used to fix the grid at the edges. TEM BP samples were stored in a $N_2$ glove box before AFM measurements. To find the same flake imaged in TEM, a built-in microscope was used to position the cantilever. To avoid damage to the cantilever and TEM grid, we avoided landing the cantilever in the holes of the carbon film and used soft tapping conditions.

**Supplementary Figures**

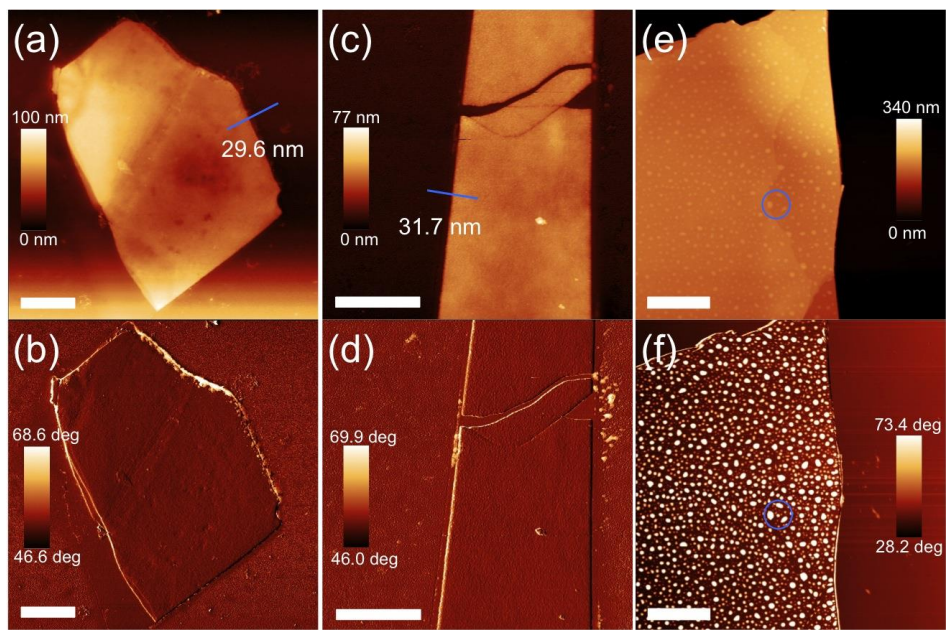

**Figure S1**. AFM images of transferred BP flakes and a degraded flake. To further verify that the wet transfer method used to prepare TEM grid samples does not cause degradation to the BP flakes, we transferred BP samples to $SiO_2$ in the same manner as the TEM grid samples. This process involves the following steps: (1) exfoliate BP onto a $SiO_2/Si$ wafer; (2) spin coat PMMA onto the sample; (3) etch $SiO_2/Si$ with aqueous 2M KOH solution; (4) rinse the PMMA/BP sample in an $H_2O$ bath; and, finally, (5) transfer the BP flakes onto another $SiO_2/Si$ wafer. **(a, c)** AFM tapping mode height images of two randomly selected flakes that were wet transferred onto $SiO_2/Si$. Both of them are ~30 nm thick. **(b, d)** Phase images of the two wet transferred flakes. No BP degradation is observable in the height and phase images. **(e)** Height and **(f)** phase images of an exfoliated BP flake stored in ambient for 3 days preceding the measurement. The blue circles indicate the same area. By comparison, degradation is evident in both the height and phase images in this case. The scale bars are 2 μm.

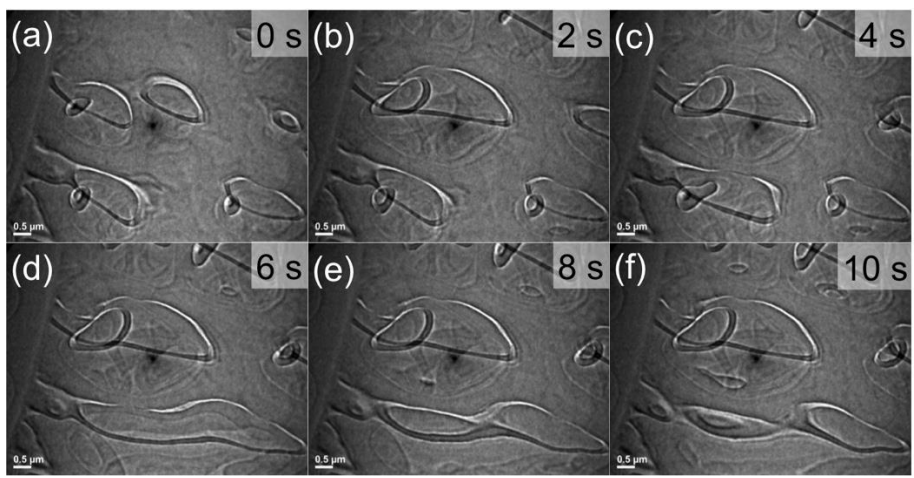

**Figure S2**. TEM images of entrapped species within ambient exposed BP. **(a-f)** TEM bright field images of entrapped liquid moving inside BP flakes. Imaging times are labeled on the images. This BP sample was stored in ambient for 5 days preceding TEM imaging.

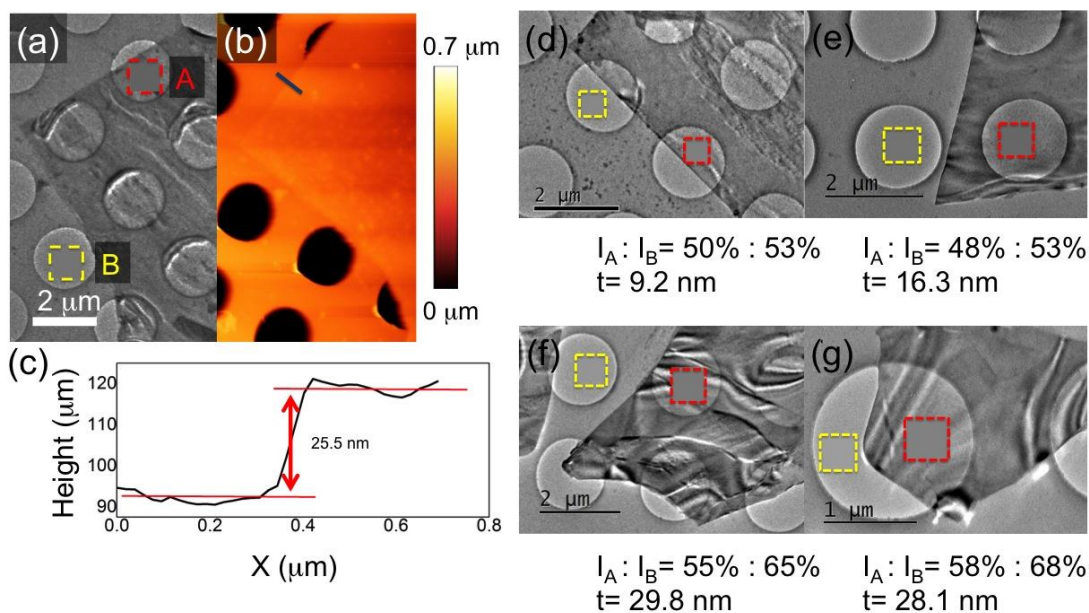

**Figure S3**. AFM and TEM imaging of the same BP flake. **(a)** TEM bright field image of an exfoliated BP flake. The accelerating voltage is 120 keV. **(b)** AFM tapping mode height image of the same flake as part (a). This flake did not undergo any *in situ* heating. **(c)** Height profile of the edge indicated in (b). The thickness of the flake is 25.5 nm. **(d-g)** TEM images of four additional flakes. To estimate the thickness range of the flakes used in this study, we calculate the thickness of the flakes based on the contrast and the known thickness of the flake in (a). Assuming that elastic scattering is dominant,[4] we can correlate the contrast with thickness by using $C = 1 - \exp(-Qt)$, where $Q$ = total elastic scattering cross section depending on atomic number, $t$ is the thickness difference, $C$ is the contrast defined as $C = (I_B - I_A)/I_A$, and $I$ is the brightness.[4] In our case, $I_A$ is the brightness of BP flakes supported on the holes of the carbon film as indicated by the red squares in Figs.

S3(a, d-g). $I_B$ is the brightness of vacuum background as indicated by the yellow squares in Figs. S3(a, d-g). The BP flake thickness values are given by $t$. Since we have the same material, $Q$ is a constant. We determine the average brightness of area A and B in Fig. S3(a) to be $I_A$ = 38% and $I_B$ = 44%, respectively. $Q$ is then determined to be $Q$ = 6.739 × 10$^{-3}$/nm. Similarly, the brightness in (d-g) is obtained, and the thickness is calculated as indicated in the figure. Hence, the thickness distribution of the BP flakes is estimated to be 10 to 40 nm.

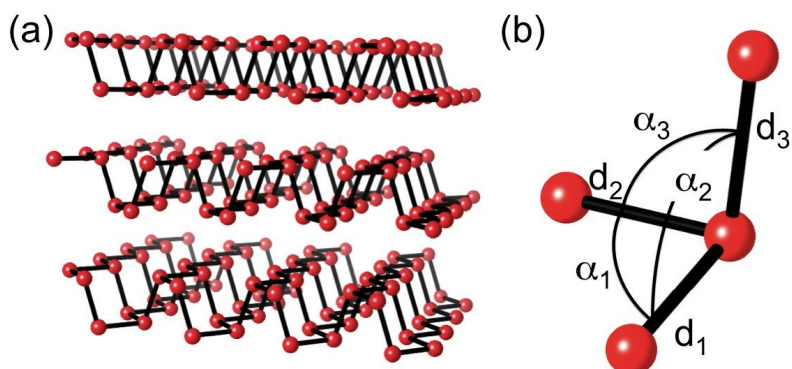

**Figure S4.** Bonding configuration for BP.[5] **(a)** Structural model for BP, showing its layered nature. **(b)** Bond lengths and bond angles for BP. BP possesses two in-plane $d_1$=$d_2$= 2.22 Å bonds and one out-of-plane $d_3$=2.24 Å bond. Bond angles are $\alpha_1$=96.34° and $\alpha_2$=$\alpha_3$=102.09°, respectively.

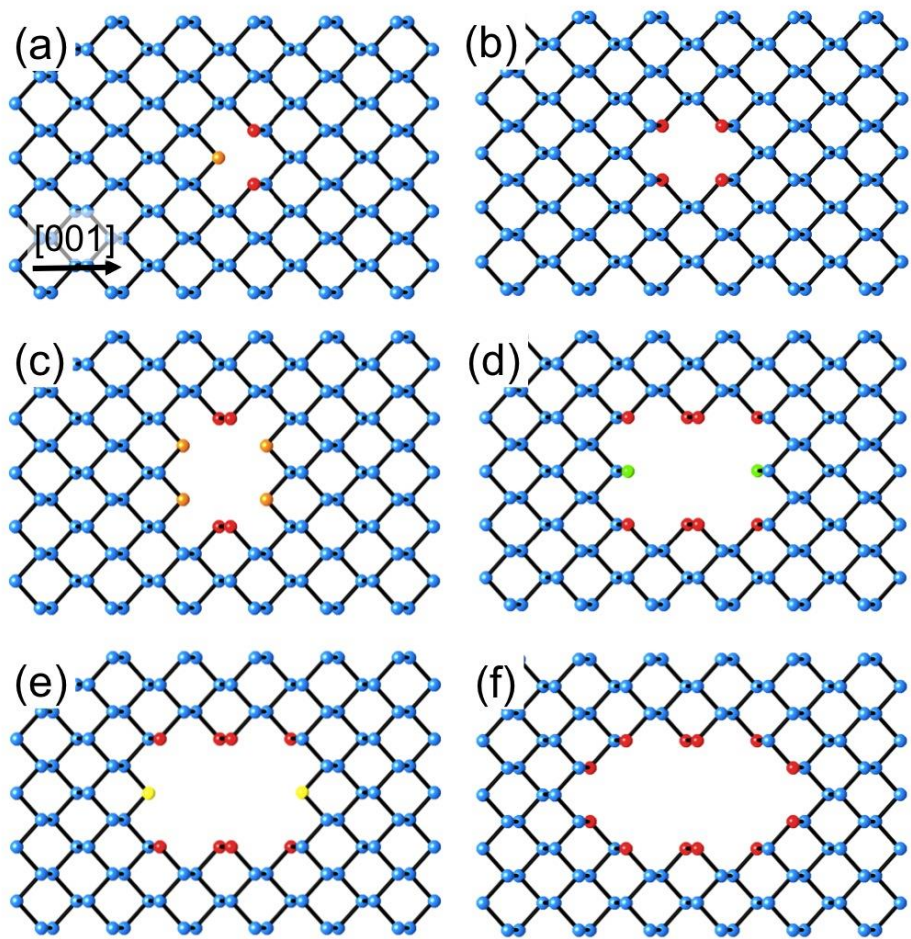

**Figure S5**. Frame-by-frame illustration of the crack growth process. **(a)** Starting point defect and **(b-f)** evolution of the crack. Based on the model described in the main manuscript, P atoms with only one bond (colored green) to the surrounding crystal desorb first, followed by P atoms with two in-plane bonds (colored yellow), and finally P atoms with one in-plane bond and one out-of-plane bond (colored red). Sequential desorption in this manner leads to preferential crack evolution along the [001] direction.

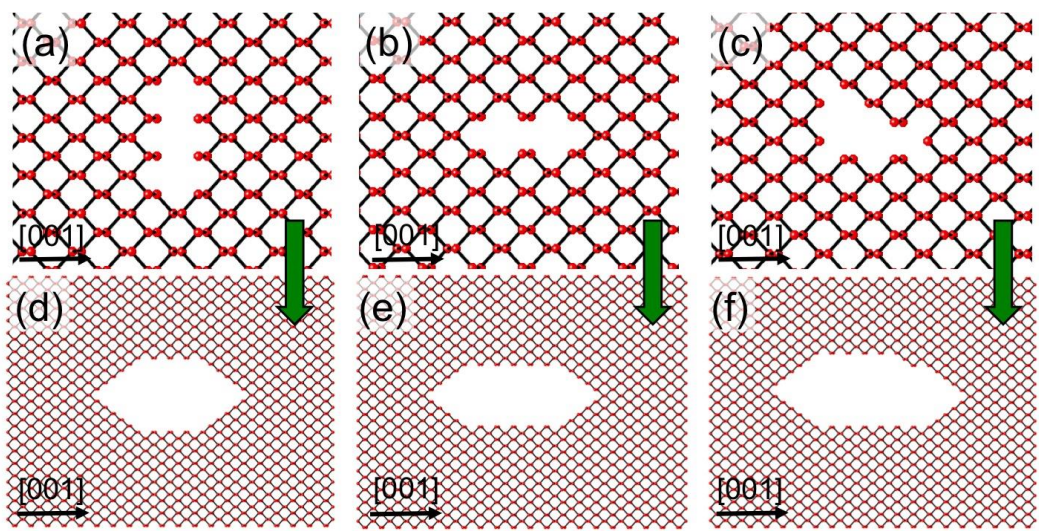

**Figure S6.** Growth of eye-shaped cracks from a series of extended vacancy defects. Example initial defects are shown in **(a-c)** including **(a)** a linear defect along the [100] direction, **(b)** a linear defect along the [001] direction, and **(c)** an extended defect with an irregular shape. **(d-f)** The growth of eye-shaped cracks from the small defects in (a-c). Based on the aforementioned order of bond breaking, all of these initial defects result in similar eye-shaped cracks along the [001] direction. For the asymmetric defect with irregular shape in (c), the eye-shaped crack will have similar asymmetry, although this asymmetry will remain at the same scale as the initial defect, *i.e.*, at the atomic scale. Thus, at larger scales (e.g. ~100 nm), this asymmetry will not be visible.

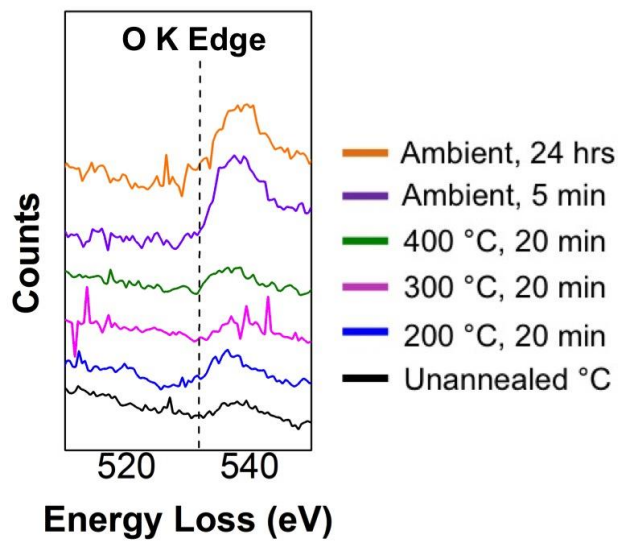

**Figure S7.** Subtracted and normalized O K edge (532 eV) of EELS spectra taken during *in situ* STEM heating and after *ex situ* ambient exposure. The oxygen peak increases significantly following ambient exposure.

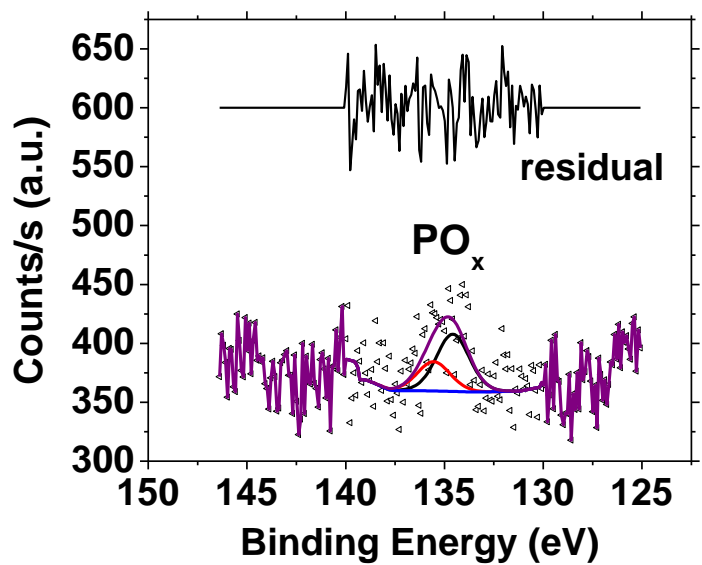

**Figure S8.** XPS spectrum of a heated BP TEM sample after *ex situ* exposure. A doublet that corresponds to oxidized phosphorus ($PO_x$) appears in the spectrum.[6] Residual from the fit offset for clarity. The individual $PO_x$ subpeaks are at 134.6 eV and 135.6 eV for $2p^{3/2}$ and $2p^{1/2}$, respectively.

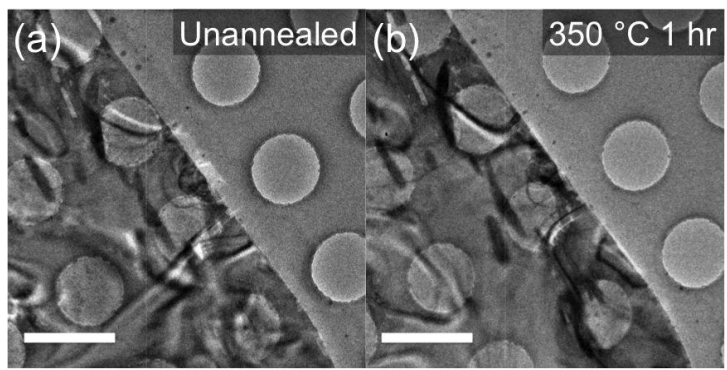

**Figure S9.** TEM images of a BP flake heated at 350 °C for 1 hour. In this experiment, we slowly increase the annealing temperature to 350 °C over 1.5 hr and then maintain the sample temperature at 350 °C for an additional 1 hr. A pristine BP flake **(a)** before and **(c)** after 1 hr annealing at 350 °C. Other than the flattening of some wrinkles, the BP flake remains intact. The scale bars are 2 μm.

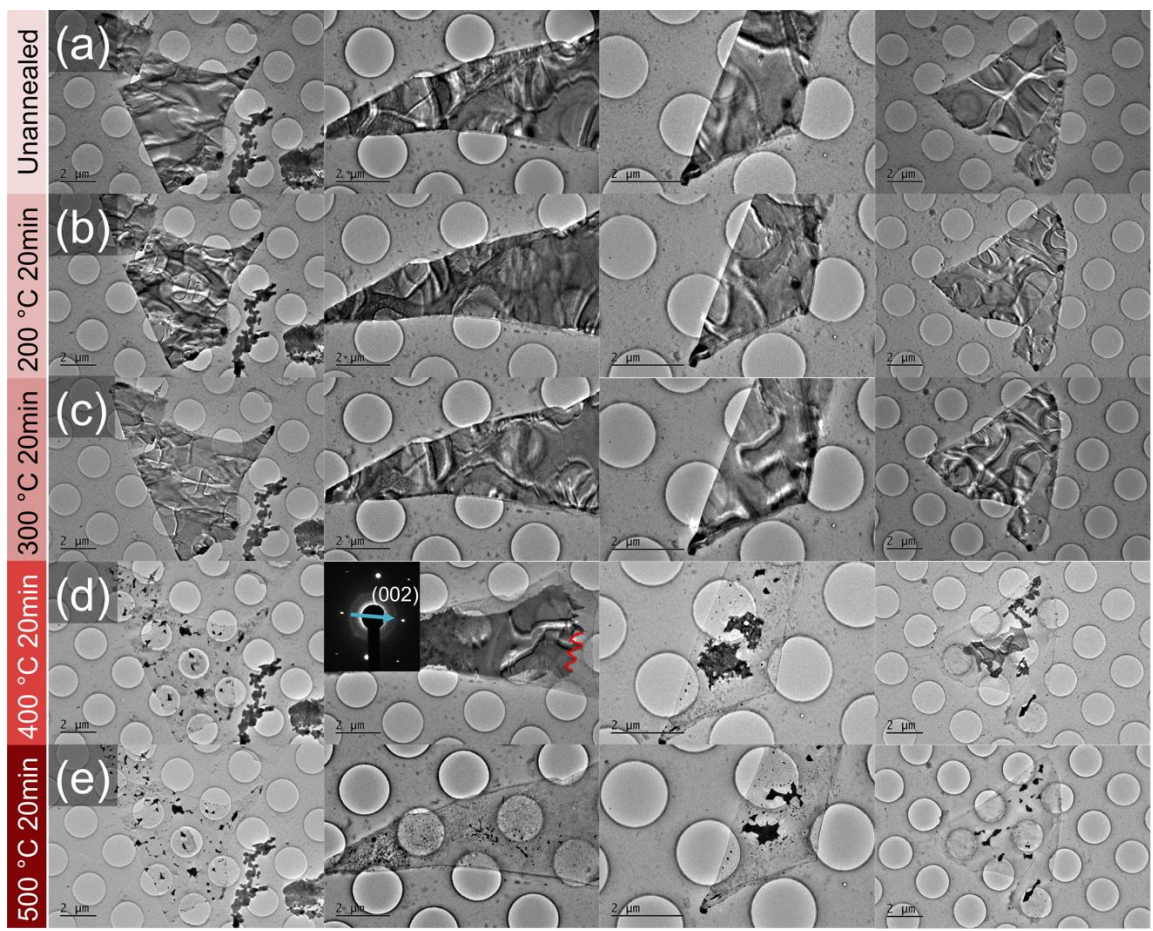

**Figure S10.** TEM bright field images of flakes heated at different temperatures. **(a-e)** Four different flakes before annealing and after annealing at 200 °C for 20 min, 300 °C for 20 min, 400 °C for 20 min, and 500 °C for 20 min, respectively. As discussed in the main manuscript, sublimation occurs around 400 °C, after which a thin skeleton remains. After 400 °C heating for 20 min, the second flake has more material left than other flakes, implying that this flake is thicker than the others. The removal of atoms in one layer enables the removal of atoms in subsequent layers. Hence, thicker flakes will require a longer time to fully sublimate. Also, from Langmuir theory,[7] the sublimation rate is constant, meaning thicker flakes take a longer time to sublimate. The inset of the second flake in (d) is the SAD pattern of the flake taken at 300 °C. The eye-shaped cracks at the edge (red line) are along the [001] direction indicated by the blue arrow in the SAD pattern, which is consistent with the main manuscript.

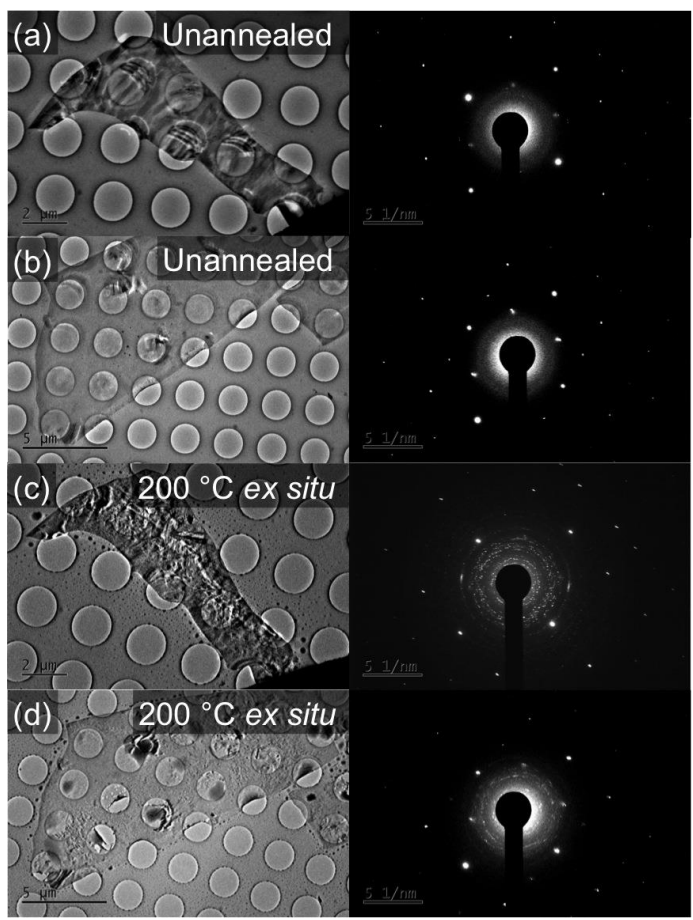

**Figure S11.** TEM and SAD imaging for flakes that underwent 200 °C *ex situ* heating. The starting temperature is 26 °C at a relative humidity of 38%. **(a, b)** Two BP flakes before heating. **(c, d)** The same flakes after 200 °C *ex situ* heating in air. The flakes are rougher, and dotted contaminants appear at the perimeter of the flakes. Amorphous rings and polycrystalline points appear in the SAD patterns.